\documentstyle[11pt,aaspp4]{article}
\begin{document}
\newcommand{\msun}{M_{\odot}}
\newcommand{\zsun}{Z_{\odot}}
\newcommand{\kms}{\, {\rm km\, s}^{-1}}
\newcommand{\cm}{\, {\rm cm}}
\newcommand{\gm}{\, {\rm g}}
\newcommand{\erg}{\, {\rm erg}}
\newcommand{\mpc}{\, {\rm Mpc}}
\newcommand{\seg}{\, {\rm s}}
\newcommand{\kev}{\, {\rm keV}}
\newcommand{\angs}{\, {\rm \AA}}
\newcommand{\hz}{\, {\rm Hz}}
\def\ion#1#2{{\rm #1\,\sc #2}}
\def\OVIII{{\ion{O}{viii}}}
\def\hi{{\ion{H}{i}}}
\def\hii{{\ion{H}{ii}}}
\def\hei{{\ion{He}{i}}}
\def\heii{{\ion{He}{ii}}}
\def\heiii{{\ion{He}{iii}}}
\def\he{\rm{He}\ }
\newcommand{\nhi}{N_{\hi}}
\newcommand{\lya}{Ly$\alpha$ }
\newcommand{\etal}{et al.\ }
\newcommand{\yr}{\, {\rm yr}}
\newcommand{\eq}{eq.\ }
\def\arcsec{''\hskip-3pt .}

\title{Soft X-ray Absorption by High-Redshift Intergalactic Helium}
\author{Jordi Miralda-Escud\'e$^{1}$}
\affil{University of Pennsylvania, Dept. of Physics and Astronomy,
David Rittenhouse Lab.,
209 S. 33rd St., Philadelphia, PA 19104}
\authoremail{jordi@llull.physics.upenn.edu}
\affil{$^{1}$ Alfred P. Sloan Fellow}

\begin{abstract}
  The Lyman alpha absorption from intergalactic, once-ionized helium (\heii)
has been measured with HST in four quasars over the last few years, in the
redshift range $2.4 < z < 3.2$. These observations have indicated that the
\heii\ reionization may not have been completed until $z\simeq 2.8$, and that
large fluctuations in the intensity of the \heii-ionizing background were
present before this epoch. The detailed history of \heii\ reionization at
higher redshifts is, however, model-dependent and difficult to determine
from these observations, because the IGM can be completely optically thick
to \lya photons when only a small fraction of the helium remains as
\heii. In addition, finding quasars in which the \heii\ \lya absorption can
be observed becomes increasingly difficult at higher redshift, owing to the
large abundance of hydrogen Lyman limit systems.

  It is pointed out here that \heii\ in the IGM should also cause detectable 
continuum absorption in the soft X-rays. The spectrum of a high-redshift
source seen behind the IGM when most of the helium was \heii\ should
recover from the \heii\ Lyman continuum absorption at an observed energy
$\sim 0.1 \kev$. Galactic absorption will generally be stronger, but not by
a large factor; the intergalactic \heii\ absorption can be detected as an
excess over the expected Galactic absorption from the 21cm HI column density.
In principle, this method allows a direct determination of the fraction of
helium that was singly ionized as a function of redshift, if the
measurement is done on a large sample of high-redshift sources over a
range of redshift.

\end{abstract}


\section{Introduction}

  The \lya absorption by intergalactic \heii\ has now been observed in
four quasars at redshifts $2.4 < z < 3.2$ (Jakobsen \etal 1994;
Davidsen \etal 1996; Reimers \etal 1997; Anderson \etal 1999; Heap
\etal 1999). While it is known that hydrogen (together with \hei) was
ionized at redshift $z \gtrsim 5$, given the presence of transmitted
flux to the blue of the hydrogen \lya wavelength in sources up to that
redshift, the double ionization of helium can take place at a lower
redshift. The reason is simple: even though there are only $\sim 8$
helium atoms for every 100 hydrogen atoms, \heii\ recombines at a rate
$5.5$ times faster than hydrogen. Consequently, if a large number of
recombinations for each helium ion take place, then the \heii\
reionization will be delayed relative to the hydrogen reionization as
long as the number of photons emitted to the intergalactic medium
(hereafter IGM) above $54.4$ eV is less than 0.44 times the number of
photons emitted above $13.6$ eV. This condition is satisfied by the
known sources of ionizing radiation (quasars and galaxies). In
practice, the mean number of recombinations of \heii\ may not be very
large: for the uniform IGM, the \heii\ recombination rate is $\sim 4$
times the Hubble rate at $z=4$, and then a somewhat softer emission
spectrum is required for a delayed \heii\ reionization. However, the
recombination rate is enhanced by the clumping factor of the ionized
gas, which increases as progressively denser gas is reionized, as
described recently in Miralda-Escud\'e, Haehnelt, \& Rees (1999;
hereafter MHR).

  The present state of the observations of intergalactic \heii\ can
be summarized as follows: at redshift $z \lesssim 2.8$, there is a
\heii\ ``\lya forest'', with a flux decrement that is consistent with
the hydrogen \lya forest flux decrement, and a background spectrum
produced by quasars, plus a possible contribution
from galaxies to increase the ratio $J_{\hi}/J_{\heii}$ of intensities
at the \hi\ and \heii\ ionization edges (Davidsen \etal 1996,
Miralda-Escud\'e \etal 1996, Bi \& Davidsen 1997, Croft \etal 1997).
At $z\gtrsim 2.8$, the \heii\ \lya spectra appear to be divided into
two different types of regions: those where transmitted flux is
still observed (with a similar $J_{\hi}/J_{\heii}$ implied for the
ionizing background), and those where the transmitted flux is
undetectable or, at least, much smaller (see Heap \etal 1999 and
references therein). The regions with greater transmitted flux occupy a
small fraction of the spectrum at $z \gtrsim 3$, and they have
typical widths of $\sim 10^3 \kms$. Some of them can be attributed
to the proximity effect of the source being observed (e.g., Hogan \etal
1997).

  These observations are not yet conclusive in clarifying the
ionization history of \heii\ in the IGM. The reason is that the optical
depth of a homogeneous medium with the mean baryon density of the
universe, where all the helium is \heii, is very large:
\begin{equation}
\tau_{0,\heii} = 1.7\times 10^3 \, {\Omega_b h \, Y \over 0.007}\,
{ H_0 (1+z)^{3/2} \over H(z) }\, \left( {1+z\over 4} \right)^{3/2} ~.
\end{equation}
Therefore, if only a fraction $\sim 10^{-3}$ of the average IGM
helium density is in the form of \heii, the flux transmission can
already be reduced to very low levels. Even when the line of sight
crosses a void with density $\rho = 0.1 \bar\rho$ (about the lowest
densities in the photoionized IGM at $z=3$ according to numerical
simulations of Cold Dark Matter models; see MHR and references
therein), the void will still be optically thick to \heii\ \lya photons
if its \heii\ fraction is greater than $\sim 0.01$. The implication is
that even a very stringent upper limit to the \heii\ \lya flux
transmission does not imply that most of the \heii\ had not yet been
reionized at the observed redshift.

  Although the observations quoted earlier could be simply interpreted
as revealing an IGM with a patchy ionization of \heii, with \heiii\
regions surrounding luminous quasars and a pure \heii\ IGM filling the
volume between the \heiii\ regions, the correct picture is probably
much more complicated. In the regions where no transmitted flux is
detected, the helium might also be doubly ionized over most of the
volume, although with a much lower ionizing background intensity
than in the regions near luminous quasars. The higher \heii\ fraction
implied in the regions with low intensity, although still much smaller
than unity, could be enough to completely absorb the \lya photons. The
presence of this low intensity, but widespread, background above the
\heii\ edge is quite plausible, as discussed in
MHR, if there is a modest contribution to the emissivity from sources
of luminosity much lower than quasars.

  A more direct observational probe to the mean fraction of \heii\ in the
IGM at different redshifts would be highly valuable to test models for
the reionization history, which is in turn important for the history of
galaxy formation owing to the strong heating of the IGM gas produced by
the \heii\ reionization (e.g., Efstathiou 1992; Miralda-Escud\'e \& Rees
1994). It is shown in this paper that this new probe may be found in the
soft X-ray continuum absorption spectra of high-redshift quasars.

  The distributed \heii\ in the IGM should cause, in addition to the
\heii\ forest absorption in \lya and the other Lyman series lines, the
continuum absorption (due to photoionization) below a rest-frame
wavelength $228$ \AA. This continuum absorption should be very large
near the edge, but the flux should recover at shorter wavelengths. We
shall see that when most of the helium in the IGM is \heii, this
recovery of the flux should occur at $\sim 0.5 \kev$, or an observed
frequency of $\sim 0.1 \kev$ at $z=4$.

\section{X-Ray Absorption by Intergalactic Helium}

  Let $F(z)$ be the mean fraction of helium in the IGM in the form of
\heii\ as a function of redshift. For the moment, we consider a
homogeneous IGM (with a uniform fraction $F(z)$), but because we shall
consider continuum absorption only, the results will be valid also for a
clumpy IGM, as long as the clumps have a large covering factor. The
continuum absorption spectrum by \heii\ on a source at redshift $z_s$ is
given by
\begin{equation}
\tau(\nu)  = \int_0^{z_s} dz\, {dl\over dz} \, n_{\he,0}\, (1+z)^3 \,
F(z)\, \sigma_{\he}[\nu(1+z)] ~,
\end{equation}
where $dl$ is the proper length element along the line of sight,
$n_{\he,0}$ is the mean primordial helium density extrapolated to the
present epoch, $\sigma_{\he}(\nu)$ is the photoionization cross section
of \heii, and $\nu$ is the observed frequency. If \heii\ is highly
ionized at low redshift, most of the contribution to $\tau(\nu)$ will be
from high redshift, where we can use $dl/dz \simeq cH_0^{-1}\,
\Omega_0^{-1/2}\, (1+z)^{-5/2}$. We also define $\Sigma_{\he}(\nu)
\equiv \sigma_{\he}(\nu)\cdot (\nu/\nu_{\he})^3$, where $\nu_{\he}$ is
the frequency at the ionization edge of \heii\ (the function
$\Sigma_{\he}$ varies only slowly with frequency and is plotted below
in Fig.\ 1). Equation (1) can be rewritten as
\begin{equation}
\tau(\nu)  = { c n_{\he,0} \over H_0 \Omega_0^{1/2} }\,
\sigma_{\he}[\nu(1+z_s)]\, \left( 1+z_s \right)^3 \,
\int_0^{z_s} { dz\, F(z) \over (1+z)^{5/2} } \,
{ \Sigma_{\he}[\nu(1+z)] \over \Sigma_{\he}[\nu (1+z_s) ] } ~,
\end{equation}
or, using $\Omega_0=0.3$, $\Omega_b h^2 = 0.019$, $Y=0.25$,
$H_0=65 \kms\mpc^{-1}$:
\begin{equation}
\tau(\nu)  = 550 \, \left( {\nu_{\he}\over \nu} \right)^3\,
{\Sigma_{\he}[\nu(1+z_s) ] \over \Sigma_{\he}(\nu_{\he}) }\,
\int_0^{z_s} {dz\, F(z) \over (1+z)^{5/2} }\,
{\Sigma_{\he}[\nu(1+z)] \over \Sigma_{\he}[\nu(1+z_s)] } ~.
\label{hzs}
\end{equation}

  As an example, we consider the simple case of a sudden and complete
reionization of \heii\ at $z=z_i$: $F=1$ at $z > z_i$ and $F=0$ at
$z < z_i$. The integral in the above equation is then well
approximated as $(z_s - z_i)/(1+z_s)^{5/2}$ (for small $z_s - z_i$).
For $z_i=3$ and $z_s = 4$, $\tau(\nu) \simeq 10 (\nu_{\he}/\nu)^3$, so
the optical depth reaches unity at an observed frequency
$\nu\simeq 0.12$ keV.

\section{Comparison to Galactic absorption}

  Any extragalactic source will also be absorbed in soft X-rays by
Galactic hydrogen and helium, with a hydrogen column density $N_{\hi}$
that can be determined from the \hi\ 21 cm emission. The absorption
optical depth is
\begin{equation}
\tau_G(\nu) = N_{\hi}\,
\left[ \sigma_{\hi}(\nu) + 0.084 \sigma_{\hei}(\nu) \right] \equiv
N_{\hi}\, \sigma_{\hi}(\nu)\, R_{\he}(\nu) ~,
\label{gals}
\end{equation}
where the ratio of helium to hydrogen atoms is $(Y/3.97)/(1-Y)=0.084$,
$\sigma_{\hi}$ and $\sigma_{\hei}$ are the photoionization cross
sections of \hi\ and \hei, and we have defined the quantity
$R_{\he}(\nu)$. At the frequencies that will be of interest to us,
$h\nu \simeq 0.2$ keV, this quantity is $R_{\he}(\nu)\simeq 3$ (this
is shown in detail below, in Fig.\ 1).

Dividing the optical depth due to the high-redshift IGM from (\ref{hzs})
by the Galactic optical depth in (\ref{gals}), we obtain the ratio to be:
\begin{equation}
{ \tau(\nu)\over \tau_G(\nu) } = {56 \over N_{\hi,20} R_{\he}(\nu)}\,
{\Sigma_{\he}[\nu(1+z_s)] \over \Sigma_{\he}(4\nu) } \,
\int_0^{z_s} {dz\, F(z) \over (1+z)^{5/2} }\,
{\Sigma_{\he}[\nu(1+z)] \over \Sigma_{\he}[\nu(1+z_s)] } ~,
\end{equation}
where $N_{\hi,20} = N_{\hi}/(10^{20}\cm^{-2})$. This fiducial \hi\
column density is around the lowest value that is usually reached at
high Galactic latitude (e.g., Laor \etal 1997).
Using the same example as in \S 2, where the integral is given by
$(z_s-z_i)/(1+z_s)^{5/2} \simeq 0.018$, we find
$\tau(\nu)/\tau_G(\nu)\simeq 1/[N_{\hi,20} \, R_{\he}(\nu)] \simeq
1/(3 N_{\hi,20})$. Thus, we conclude that the absorption by
high-redshift intergalactic helium should be smaller than the
Galactic absorption only by a factor $\sim 3$ when the Galactic
column density has the lowest value normally found in sources,
$N_{\hi}\simeq 10^{20}\cm^{-2}$.

  Laor \etal (1997) measured the Galactic column densities in a sample
of 23 low-redshift quasars from the absorption in soft X-rays. Comparing
the column densities they derived with those measured from the 21 cm
emission, they show that there are no significant differences within
observational error. In particular, the two quasars with the smallest
error (10\%) in the column density determined from the soft X-ray
spectrum by Laor \etal also agree with the 21 cm column density
(see their Fig.\ 2). Provided that the Galactic absorption
can be subtracted from the measured X-ray absorption on a high-redshift
quasar, given the column density measured from 21cm emission,
it should be possible to detect the continuum absorption of
intergalactic helium in the soft X-rays.

  A natural question to ask here is if the shape of the absorption can
be used to distinguish Galactic absorption from the high-redshift \heii\
absorption. Unfortunately, the shape is almost identical in the two
cases, as we shall now see. In Figure 1, the solid line is the hydrogen
cross section $\sigma_{\hi}(\nu)$, multiplied by $(\nu/\nu_{\hi})^3$,
and the dashed line is $[\sigma_{\hi}(\nu) + n_{\he}/n_{\rm H}\,
\sigma_{\hei}(\nu) ]\, (\nu/\nu_{\hi})^3$ [so the ratio of the dashed
line to the solid line is $R_{\he}(\nu)$, defined in \eq (\ref{gals})].
We have used the exact analytical expression for $\sigma_{\hi}(\nu)$
(e.g., Spitzer 1978), and the fit of Verner \etal (1996) for
$\sigma_{\hei}(\nu)$. The dashed line is therefore the expected shape
of the Galactic absorption, except for the fact that some of the
ionized gas in the Galaxy may not have the helium doubly ionized, and
there could then be some additional \heii\ absorption at $z=0$. This is
shown by the dotted line, equal to $\sigma_{\he}(\nu)\,
(n_{\he}/n_{\rm H}) \, (\nu/\nu_{\hi})^3 $, and we see that it has almost
the same slope as the dashed line at energies $\sim 0.15$ keV.

  The intergalactic \heii\ absorption should have the same shape as the
solid line, at redshift $z=3$ (because the \heii\ cross section is
identical to the \hi\ cross section shifted to a frequency 4 times
higher). Notice from equation (4) that the frequency dependence of
$\tau(\nu)$ is basically the same as the redshifted cross section
$\sigma_{\he}(\nu)$. Thus, we see from Fig. 1 that the high redshift
absorption has a slightly steeper cross section than the Galactic one.
In the range from $0.1$ keV to $0.2$ keV, the Galactic cross section
is $\sigma \propto \nu^{-3.0}$, and the intergalactic helium cross
section is $\sigma_{\he}\propto \nu^{-3.15}$. In practice, it will be
extremely difficult to reach the high signal-to-noise required to
distinguish between these two slopes; moreover, the intrinsic emission
spectrum of the source introduces an additional uncertainty. Therefore,
the high-redshift helium absorption can probably be detected only as an
excess of absorption over that expected from the Galactic column
density derived from the 21 cm emission.

\section{A Reionization Model}

  In any realistic model, the reionization of all the \heii\ in the
universe will take place over a substantial length of time, of the
order of a Hubble time. In the reionization model proposed in MHR, the
low-density gas is ionized earlier and reionization advances outside-in.
The stage of reionization depends at every redshift on the gas
overdensity $\Delta_{\he}$ up to which the helium is mostly ionized to
\heiii. Using results of numerical simulations for the density
distribution and the photon mean free path in the IGM, the model in MHR
predicts the \heii\ \lya flux decrement and the fraction of gas at
overdensities greater than $\Delta_{\he}$, assumed to be still in the
form of \heii. The observed flux decrement of $\sim 0.7$ at $z=2.6$
(Davidsen \etal 1996) requires $\Delta_{\he}\simeq 70$, and the fraction
of \heii\ $F\simeq 0.2$ (see Figs. 2 and 9 in MHR). At the slightly
higher redshift $z\simeq 3.1$, a flux decrement as high as $0.99$
(see Heap \etal 1999) requires $\Delta_{\he} \simeq 12$, and the
fraction of gas at greater overdensities increases only to
$F \simeq 0.3$, showing how a small increase in \heii\ ionization
can result in a dramatic decline of the transmitted flux due to the
effect that the most underdense voids rapidly become optically thick
to the \heii\ \lya photons.

  As an example of a reasonable model matching the above conditions,
where the \heii\ reionization takes place over an extended period of
time, we consider the simple case $F=[(1+z)/7]^{5/2}$, where the
first \heii\ sources would turn on at $z=6$. For this case, equation
(6) yields (neglecting the ratios of the function $\Sigma_{\he}$)
$\tau(\nu)/\tau_G(\nu) \simeq 0.43\, z_s/N_{\hi,20}/R_{\he}(\nu)$
(for $z_s < 6$).

  It must be born in mind here that whatever contribution arises from
low redshifts to the integral in equation (6) should originate mostly in
dense systems, where the helium is self-shielded and remains in the
form of \heii\ (these are the observed Lyman limit and damped absorption
systems). When the mean free path between these absorption systems is
not much smaller than the Hubble length (i.e., when the absorbers do not
have a large covering factor), the \heii\ column density on a particular
line of sight will no longer be equal to the mean value, but will have
large fluctuations: it will most often be smaller than the mean value,
but it will exceed this mean value when a rare, strong absorption system
is intersected. The presence of these rare absorption systems, which
should dominate the total \heii\ content of the universe at low redshift
only, can in principle be determined independently for every line of
sight from observations of the \hi\ absorption spectrum. In practice,
though, the Lyman limit absorption from higher redshift systems, and the
\heii\ \lya absorption itself, will prevent the determination of the \hi\
column density of any low-redshift absorber in a high-redshift source.
The absence of strong metal lines can still be a good indication that
a strong \hi\ absorber is not present.

  In order for a low-redshift absorber to cause soft X-ray absorption
comparable to that of the high-redshift IGM, it must have a column
density $\nhi\gtrsim 10^{19.5} (1+z)^3 \cm^{-2}$. Only $\sim$ 20\% of
the lines of sight have an absorber with this strength (e.g.,
Storrie-Lombardi, Irwin, \& McMahon 1996). Nevertheless,
the fluctuation in the low-redshift absorption will be an additional
source of uncertainty (especially if the low-redshift \hi\ \lya
absorption cannot be measured), implying that the excess soft X-ray
absorption will need to be detected in several sources before one can
be sure that the effect of the helium in the low-density IGM at high
redshift has been detected.

  The fluctuations in the \heii column density due to high-redshift
absorbers of high column density are less important, because these
absorbers do not contain a large fraction of the baryons. For example,
at $z=3$ we need an absorber with $\nhi \sim 10^{21}\cm^{-2}$ to
produce an absorption similar to the IGM. These absorbers are rare,
and they can also be readily identified in the \hi\ \lya spectrum
when their redshift is not much lower than the source redshift.

\section{Conclusions}

  Soft X-ray absorption can be a new powerful tool to measure the
fraction of helium in the form of \heii\ as a function of redshift.
This direct determination of the \heii\ fraction provides a
straightforward test of any model of reionization based on the
observed emitting sources (quasars and galaxies). Although the
observations of the \heii\ \lya absorption spectra provide a much
greater wealth of information, their interpretation is more
complicated due to the complexities introduced by the highly
inhomogeneous IGM and the fluctuating intensity of the ionizing
background.

  The main challenge in detecting this effect will be to find enough
high-redshift, X-ray bright sources, and to accurately measure the
21 cm \hi\ column density to subtract the Galactic contribution to
the soft X-ray absorption. A systematic uncertainty that
will be faced is the possible existence of some Galactic \heii\ along
the line of sight, which can also produce excess soft X-ray absorption
not accounted for by the Galactic \hi\ column density. In addition,
extragalactic damped \lya absorbers at low redshifts can also make a
significant contribution, which may be difficult to correct for in
high-redshift sources. These systematic uncertainties can be put
under control once the effect is measured in many sources over a
wide range of redshift.

\acknowledgements

  I am grateful to Andy Fabian for encouraging discussions on the
possibilities for observing the effect proposed here.

\newpage

\begin{figure}
\centerline {
\epsfxsize=4.7truein
\epsfbox[70 132 545 640]{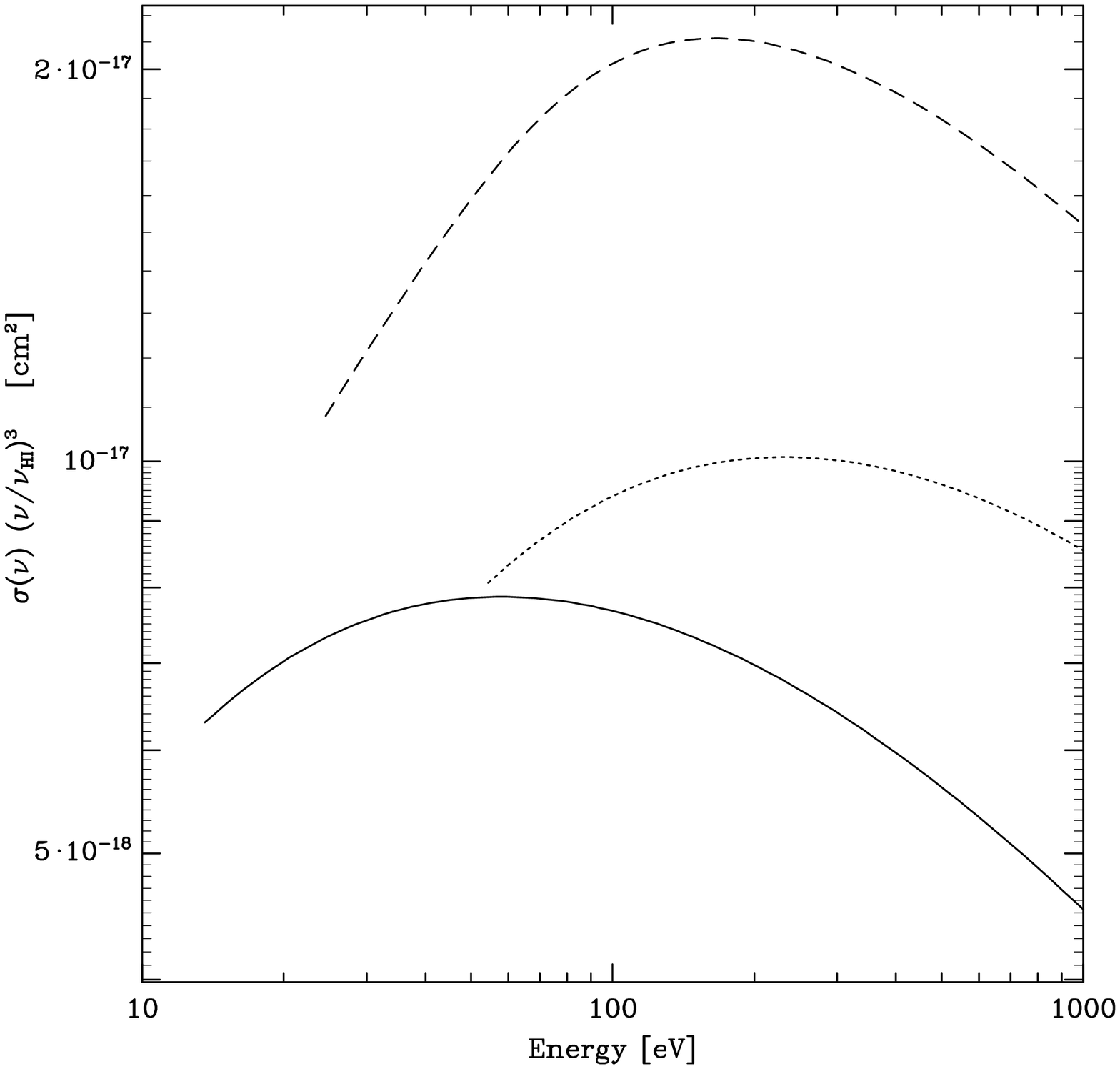}
}
\caption{Solid line is the \hi\ cross section times $(\nu/\nu_{\hi})^3$.
Dashed line is equal to solid line times $R_{\he}(\nu)$ [in \eq (5)],
giving the effective Galactic cross section including \hei. Dotted line
is the \heii\ cross section, times $(n_{\he}/n_{\rm H})
(\nu/\nu_{\hi})^3$.
}
\end{figure}

\end{document}